\documentclass[twocolumn,floats,floatfix,aps,pra]{revtex4-2}
\usepackage{amsfonts,amssymb,amsmath}
\usepackage{color,calc}
\usepackage[dvips]{graphicx}
\usepackage{bm}
\usepackage{array}
\usepackage{hyperref}
\usepackage{physics}
\usepackage{arydshln}

\def\be{ \begin{equation} }
\def\beal{ \begin{align} }
\def\ee{ \end{equation} }
\def\bea{ \begin{eqnarray} }
\def\eea{ \end{eqnarray} }
\def\bse{ \begin{subequations} }
\def\ese{ \end{subequations} }
\def\ba{ \begin{array} }
\def\ea{ \end{array} }
\def\bwt{ \begin{widetext} }
\def\ewt{ \end{widetext} }

\def\1{\scalebox{1.5}a}
\def\2{\scalebox{1.5}b}
\def\3{\scalebox{1.5}c}

\def\b{b}

\def\k{k}

\def\S{\mathbf{S}}
\def\H{\mathbf{H}}
\def\U{\mathbf{U}}

\def\min{\text{min}}
\def\max{\text{max}}
\def\V{\mathbf{V}}

\def\Omg{\mathbf{\Omega}}
\def\wtl{\widetilde}

\def\k{\kappa}
\def\k{k}

\def\rho{\rho}
\def\eta{\eta}
\def\D{\mathbf{D}}

\def\Re{\textrm{Re}}
\def\Im{\textrm{Im}}

\def\K{M}
\def\l{l}
\def\r{r}
\def\R{{\bf R}}
\def\A{\mathbf{A}}
\def\B{\mathbf{B}}

\begin{document}

\title{Coherent interaction of multistate quantum systems possessing the Majorana and Morris-Shore dynamic symmetries with pulse trains}

\author{Stancho G. Stanchev and Nikolay V. Vitanov}

\affiliation{Department of Physics, St Kliment Ohridski University of Sofia, 5 James Bourchier blvd, 1164 Sofia, Bulgaria}

\date{\today }

\begin{abstract}
We present exact analytic formulae which describe the interaction of multistate quantum systems possessing the Majorana and Morris-Shore dynamic symmetries with a train of pulses.
The pulse train field can be viewed as repeated interactions of the quantum system with the same field and hence the overall propagator is expressed as the matrix power of the single-pulse propagator.
Because of the Majorana and Morris-Shore symmetries the multistate dynamics is characterised by intrinsic two-state features, described by one or more pairs of complex-valued Cayley-Klein parameters.
This facilitates the derivation of explicit formulae linking the single-step and multi-step propagators.
The availability of such analytic relations opens the prospects for a variety of applications, e.g., analytic description of coherent pulse train interactions, or coherent amplification of quantum gate errors for accurate quantum gate tomography for ensembles of qubits, qutrits and generally qudits.
\end{abstract}

\maketitle


\section{Introduction}

Quantum systems with multiple states, as encountered in real physical systems, are generally difficult to treat analytically due to their prohibitively complex dynamics \cite{Shore1990, Shore2011}. 
Yet, some of them allow for such a treatment thanks to an intrinsic two-state behavior.
One example is the situation when all but two of the states are far off resonance with any of the driving fields; then adiabatic elimination of all these states leaves us with an effective two-state system and the effect of the other states show up as ac Stark shifts in the detuning \cite{Shore1990, Shore2011}.
Interestingly, a modified adiabatic elimination can be performed even when the far-off-resonance condition is alleviated to a near-resonance one \cite{Torosov2012}.

Another example is found in systems with the Morris-Shore symmetry \cite{Morris1983}, in which all states can be grouped into two manifolds, such that interactions are allowed between states of different manifolds only, but not within the same manifold. 
All couplings must share the same time dependence.
Moreover, all states within the same manifold should be degenerate (in the rotating-wave approximation), but there could be a nonzero detuning between states from different manifolds; however, it should be the same for all couplings.
The Morris-Shore transformation casts such multistate systems into a set of independent two-state systems and a number of decoupled (dark) single states.
Such linkages naturally emerge, e.g., in the interaction of two degenerate atomic levels with an elliptically polarised laser field \cite{Shore1990,Shore2011,Vitanov2000, Vitanov2003, Kis2004, Kyoseva2008, Kim2015, Kirova2017, Finkelstein2019}.
The Morris-Shore transformation has been generalized to an arbitrary many manifolds of degenerate states \cite{Rangelov2006}, and its extensions and applications have been reviewed by Shore \cite{Shore2013}.
Recently, this transformation has been generalized to unequal detunings \cite{Zlatanov2020} and different time dependences of the couplings \cite{Zlatanov2022}.

A third example of reducible multistate dynamics is found in systems with the Majorana SU(2) symmetry \cite{Majorana1932}. 
It arises, e.g., in (radio-frequency) transitions between the magnetic sublevels of a level with a definite angular momentum, such as in Bose-Einstein output couplers \cite{Cook1979,Vitanov1997}.
It naturally emerges also in Raman systems subjected to combined external laser field and static magnetic field \cite{Randall2018}. 

While the adiabatic elimination is an approximate method, the other two methods present, in principle, exact reduction of the multistate systems to one or more two-state systems.
This reduction allows one to use the two-state quantum control methods to design similar methods in multistate systems \cite{Vitanov1997, Allen1975, Randall2018}.

In this paper, we use these analogies between two-state and multistate systems in order to describe analytically the interaction of a multistate system with either the Majorana or Morris-Shore symmetry with a train of identical pulses, viewed as a multi-pass interaction of the system with the same field.
We derive explicit analytic formulae which express the overall multi-pass propagator in terms of the single-pulse propagator. 
In addition to the obvious application of using these results in order to develop analytic models of multi-pulse excitation of multistate systems, they present the framework for the development of precise quantum tomography of such systems by coherent error amplification due to the repeated application of the same quantum gate. 

The paper is organized as follows.
Section \ref{Sec:Rep} define the problems and the framework of their treatment.
Section \ref{Sec:Majorana} presents the multi-pass interaction of systems with the Majorana symmetry, with an explicit example for a three-state system.
Section \ref{Sec:Morris-Shore} discusses the multi-pass interaction of systems with the Morris-Shore symmetry, with explicit examples for the general multipod systems and their simplest and most important cases of the Raman three-level system and the tripod system.
Finally, Section \ref{Sec:Conclusion} presents a summary of the results.


\section{Case studies }\label{Sec:Rep}

We consider a coherently driven quantum system with $\K$ states driven by a train of $N$ identical pulses of duration $T$ each.
Its evolution is governed by the time-dependent Schrodinger equation \cite{Shore1990, Shore2011} ($\hbar=1$),
\be
i \frac{d}{dt} \Psi(t) = \H(t) \Psi(t). 
\ee
The evolution of the state vector $\Psi(t)$ can be described by the propagator $\U(t,t_0)$ as
\be
\Psi(t)=\U(t,t_0) \Psi(t_0).
\ee
Without loss of generality, we take $t_0=0$.
If the Hamiltonian $\H(t)$ has the same form for each time interval $[(n-1)T,nT]$ ($n=1,2,\ldots,N$), during which the $n$-th pulse acts, then the propagator generated by each pulse will be the same, i.e. $\U((n-1)T,nT) = \U(T,0)$.
Then
\be\label{UNT}
\U(NT,0)\equiv[\U(T,0)]^N.
\ee

We assume that the single-pulse propagator $\U(T,0)$ is known, and we wish to find the $N$-pulse propagator $\U(NT,0)$ in terms of the parameters of $\U(T,0)$.
This problem has been explicitly resolved for a two-state system \cite{Vitanov1995} and we will use and build up on these results here.
%

The dynamics of a two-state quantum system is governed by the Hamiltonian
\be\label{H2}
\H_{2}(t) = \frac{1}{2}\left[\begin{array}{cc} -\Delta(t)  & \Omega(t) \\ \Omega^*(t) &\Delta(t) \end{array}\right],
\ee
where the detuning $\Delta(t)$ and the Rabi frequency $\Omega(t)$ are arbitrary functions of time.
Because we have chosen to express the Hamiltonian in the traceless form \eqref{H2}, the propagator has the SU(2) dynamic symmetry,
\be\label{U2}
\U_{2}= \left[\begin{array}{cc} a  & b \\ -b^* & a^*  \end{array}\right],
\ee
where $a$ and $b$ are two complex-valued Cayley-Klein parameters, with $|a|^2 +|b|^2=1$ and hence $\det \U_{2} = 1$.
It has been proved in ~\cite{Vitanov1995} 
 that the $N$-th power of any SU(2) propagator reads
\be\label{U2N}
\U_{2}^N = \left[\begin{array}{cc} a_N  & b_N \\ -b_N^* & a_N^*  \end{array}\right] ,
\ee
where
\bse \label{aNbN}
\begin{align}
a_N &= \cos (N\theta) + i\Im\,a\,\frac{\sin (N\theta)}{\sin \theta},\\
b_N &= b\, \frac{\sin (N\theta)}{\sin \theta},\\
\cos\theta &= \Re\,a . \label{theta}
\end{align}
\ese

The relations \eqref{aNbN}  make it possible to find out how a pulse train affects the two-state system if we know the single-pulse action.
There exist a number of analytic single-pulse solutions \cite{Rosen1932, Landau1932, Zener1932, Stueckelberg1932, Majorana1932, Allen1975, Hioe1984, Silver1985}, which can be generalized using the formulae above to analytic multiple-pulse solutions \cite{Vitanov1995}.

On the other hand, these relations allow one to precisely find out the action of a single pulse from the action of a train of pulses.
This is important, for instance, for measuring small deviations from a desired single-pulse action, e.g., for characterizing a high-fidelity quantum gate \cite{Vitanov2020,Vitanov2021} for which the admissible error is of the order of $10^{-3}$ or less.

Hitherto, such single-pass to multi-pass relations have been known for a two-state system only.
Here we extend these results to multistate systems possessing the Majorana or Morris-Shore symmetries, which are reducible to one or more two-state systems.

\begin{itemize}
\item
\textbf{Majorana symmetry.}
It  fulfils the requirements of  Majorana decomposition~\cite{Majorana1932}, having a reducible dynamic symmetry from SU($\K$) to SU(2)~\cite{Majorana1932,BlRb1945,Hioe1987,Randall2018}.\\ 

\item
\textbf{Morris-Shore (MS) symmetry.}
 The MS Hamiltonian fulfils the requirements of the MS decomposition~\cite {Morris1983,Rangelov2006,Kyoseva2006,Zlatanov2020},
 for which the quantum system is composed by a set of $L$ ground degenerate states and a set of $M$ exited degenerate states. 
 Couplings exist between states from different sets only, but not within the same set. 
 Such a system is reducible to a set of $M$ independent two-state systems and $M-L$ decoupled (dark) states. 

\end{itemize}

Below we consider first the multiple interactions of multistate systems with the Majorana symmetry, and then systems with the Morris-Shore symmetry.


\section{Systems with the Majorana symmetry\label{Sec:Majorana}}

\subsection{Majorana propagator}\label{Sec:MajoranaSingle}

The Majorana decomposition has been presented in several papers, see eg Refs.~\cite{Majorana1932,BlRb1945,Hioe1987,Randall2018}.
It stems from 
 the rotation group theory for the angular momentum. 
The Hamiltonian has the tridiagonal form~\cite{Hioe1987}
\be\label{HMajorana}
\H_{\K}\! =\! \left[\!
\begin{array}{cccccc}
H_{11}&H_{12}&0&\cdots& 0 &0 \\
H_{21}&H_{22}&H_{23}&\cdots &0 &0\\
0 & H_{32} & H_{33}& \cdots &0 &0\\
\vdots& \vdots & \vdots & \ddots & \vdots & \vdots\\
0& 0& 0 & \cdots & H_{\K-1,\K-1}& H_{\K-1,\K}\\
0&0&0& \cdots & H_{\K,\K-1} &H_{\K\K}
\end{array}
\!\right]\!,
\ee
where the nonzero matrix elements read
\bse
\begin{align}
 H_{\k\k}(t)&= 
  \Big(\k-\frac{\K+1}{2}\Big)\Delta(t) ,\\
 &(\k=1,2,...,\K), \notag \\
 H_{\k+1,\k}(t) &= H^*_{\k,\k+1}(t)=\frac{1}{2}\sqrt{\k(\K-\k)}\,\Omega(t),\\ 
 &(\k=1,2,...,\K-1), \notag
\end{align}
\ese
In terms of the angular momentum quantum numbers $j$ and $m$, we have the relations $M=2j+1$ and $k=j+1-m$. 

The matrix elements of the propagator $\U_{\K}$ are \cite{NeuWig1928,Wig1931gr,Hioe1987}
\begin{align}\label{Ukl}
U_{\k\l}&=\sum_\r \frac{\sqrt{(\k-1)! (\l-1)! (\K-\k)! (\K-\l)!}}{(\l-1-\r)! (\K-\k-\r)! (\r -\l+\k)!\r!}\notag \\
&\times a^{\K-\k-\r} (a^*)^{\l-1-\r} b^{\r } (-b^*)^{\r -\l+\k},
\end{align}
where $\r$ runs from $\r_\min$ to $\r_\max$, with
\bse \label{rmin-rmax}
\begin{align}
\r_{\min}&= \min[0,\k+\l+1-\K],\\
\r_{\max}&= \max[\k-1,\l-1] .
\end{align}
\ese
For $\K=2$ states, the matrix is the one shown already in Eq.~\eqref{U2}. 
The propagator for  $\K=3$ and $\K=4$ states reads
\bse\label{Mexample}
\begin{align}\label{U3Maj}
\U_{3} &= \left[\begin{array}{ccc}
 a^2 & a b\sqrt{2} & b^2 \\
 - a b^*\sqrt{2} & |a|^2-|b|^2 & b a^*\sqrt{2} \\
 b^{*2} & -a^* b^*\sqrt{2} & a^{*2} \\
\end{array}\right],
\\
\U_{4}\! &= \!\!\left[\!\! \begin{array}{cccc}
a^3 & a^2 b \sqrt{3} & ab^2\sqrt{3} & b^3 \\
 - a^2 b^{*}\sqrt{3} & \!a (|a|^2-2|b|^2) &  \!b(2|a|^2-|b|^2)& \!b^2 a^*\sqrt{3} \\
 a b^{*2}\sqrt{3}  & \!b^{*} (|b|^2-2|a|^2) & \!a^{*}(|a|^2-2|b|^2) & \! b a^{*2}\sqrt{3} \\
-b^{*3} & a^{*}b^{*2} \sqrt{3} & -a^{*2}b^* \sqrt{3} & a^{*3} \\
\end{array}\!\!\right]\!\!.
\end{align}
\ese

\subsection{Multi-pass Majorana Propagator \label{Sec:MajoranaRep}} 

We shall derive the Majorana propagator of an $M$-state system after $N$ repetitions of the interaction described by the Hamiltonian $\H_M$ of Eq.~\eqref{HMajorana} in two alternative manners.
First we shall use the analogy of $M$-state to two-state dynamics and then we shall derive it by diagonalization of the propagator $\U_M$.

\subsubsection{First approach}

In the {first} approach, it is important to note that the parameters $a$ and $b$, parameterizing the propagator $\U_M$, are the same as the parameters in $\U_2$ of Eq.~\eqref{U2}.
In other words, if the Hamiltonian $\H_2$ of Eq.~\eqref{H2} generates the propagator $\U_2$ of Eq.~\eqref{U2}, then the Hamiltonian $\H_M$ of Eq.~\eqref{HMajorana} generates the propagator $\U_M$ with the matrix elements of Eq.~\eqref{Ukl}.
The implication is that we can find the propagator of the $M$-state system with the Majorana symmetry in two equivalent manners: (i) solve the Schr\"odinger equation with the $M$-state Hamiltonian $\H_M$, or (ii) solve the two-state problem with the Hamiltonian $\H_2$ and use Eq.~\eqref{Ukl} to find the propagator $\U_M$.
In either cases, the solutions should be identical.
In this sense, we say that the Majorana symmetry admits reduction of the $M$-state system to an effective two-state system.

Now consider a sequence of multiple pulses, each generating the same propagator $\U_N$.
By the same reasons, given in the preceding paragraph, the multi-pulse propagator for the $M$-state system, i.e. $\U_M^N$ can be calculated using the one for the two-state system, $\U_2^N$.
We thereby conclude that the matrix elements of the $N$-pulse propagator for the $M$-state Majorana system has the same form as the single-pulse propagator elements of Eq.~\eqref{Ukl}, 
\begin{align}\label{Ukl-N}
U_{\k\l} &= \sum_\r \frac{\sqrt{(\k-1)! (\l-1)! (\K-\k)! (\K-\l)!}}{(\l-1-\r)! (\K-\k-\r)! (\r -\l+\k)!\r!}\notag \\
&\times a_N^{\K-\k-\r} (a_N^*)^{\l-1-\r} b_N^{\r } (-b_N^*)^{\r -\l+\k},
\end{align}
with $a_N$ and $b_N$ given by Eqs.~\eqref{aNbN}, and $r$ runs from $r_\min$ to $r_\max$ given by Eq.~\eqref{rmin-rmax}.

\subsubsection{Second approach}

In the second approach, we find the multi-pulse propagator $\U_M^N$ by diagonalizing the single-pulse one $\U_{\K}$ (we drop hereafter the subscript $M$ for simplicity),
\be
\V^\dagger \U \V = \D ,
\label{diagonal}
\ee
and hence 
\be
 \U = \V \D \V^\dagger.
 \label{diagonal-inverse}
\ee
According to the 3D rotation group theory ~\cite{NeuWig1928}, the diagonal matrix $\D_{\K}$ has the form
\be\label{diag}
\D =
\left[ \begin{array}{ccccc}
   e^{-i(\K-1)\theta}&0 & \cdots &0& 0 \\
0&e^{-i(\K-3)\theta}& \cdots &0& 0\\
 \vdots& \vdots & \ddots & \vdots&\vdots \\
0&0&\cdots &e^{i(\K-3)\theta}&0\\
  0 &0& \cdots& 0& e^{i(\K-1)\theta}
\end{array}\right] ,
\ee
where the phase factors are the eigenvalues of $\U$ and $\theta$ is defined by Eq.~\eqref{theta}.
The diagonalizating matrix $\V$ is composed by the eigenvectors of $\U$. 
One can show that, quite remarkably, it has the same form (i.e. possesses the Majorana symmetry) as the propagator $\U$ of Eq.~\eqref{Ukl}, but with different CK parameters $u$ and $v$, instead of $a$ and $b$,
\begin{align}\label{Vkl}
V_{\k\l} &= 
\sum_\r
\frac{\sqrt{(\k-1)! (\l-1)! (\K-\k)! (\K-\l)!}}{(\l-1-\r)! (\K-\k-\r)! (\r -\l+\k)!\r!} \notag \\
&\times u^{\K-\k-\r} (u^*)^{\l-1-\r} v^{\r } (-v^*)^{\r -\l+\k} ,
\end{align}
with $|u|^2+|v|^2=1$. 
The relations between the parameters $u$ and $v$ of the diagonalizing matrix and the parameters $a$ and $b$ of the propagator can be found as follows.

Looking at the elements \eqref{Ukl} of the propagator $\U$, exemplified for $\K=3$ and 4 states in Eqs.~\eqref{Mexample}, it is easy to notice a prominent feature: 
the elements on the top row are the square roots of the terms in the expansion of $(a^2+b^2)^{M-1}$, ie
\be
U_{1l} = \binom {M-1}{l-1}^{\frac12} \, a^{M-l} b^{l-1}.
\ee
Indeed, due to the factorials in the denominator and the fact that $1/(-n)! = 0$ for any positive integer $n$, for $k=1$ the only nonzero contribution in the sum is for $r=l-1$. 
Similarly, the elements on the bottom row (for $k=M$) are the square roots of the terms in the expansion of $(a^{*2}-b^{*2})^{M-1}$, viz.
\be
U_{Ml} = \binom {M-1}{l-1}^{\frac12} \, (a^*)^{l-1} (-b^*)^{M-l}.
\ee
because the only nonzero contribution to the sum comes from $r=0$.
In particular, in the corners we have
\be\label{U-corners}
\begin{array}{lll}
U_{11} = a^{M-1}, & & U_{1M} = b^{M-1}, \\
U_{M1} = (-b^*)^{M-1}, & & U_{MM} = (a^*)^{M-1}.
\end{array}
\ee

Let us now assume that the diagonalizing matrix $\V$ is composed of the matrix elements of Eq.~\eqref{Vkl}.
As for $\U$, the elements on the top row are the square roots of the terms in the expansion of $(u^2+v^2)^{M-1}$, ie
\be
V_{1l} = \binom {M-1}{l-1}^{\frac12} \, u^{M-l} v^{l-1},
\ee
and the elements on the bottom row (for $k=M$) are the square roots of the terms in the expansion of $(u^{*2}-v^{*2})^{M-1}$, viz.
\be
V_{Ml} = \binom {M-1}{l-1}^{\frac12} \, (u^*)^{l-1} (-v^*)^{M-l},
\ee
because the only nonzero contribution to the sum comes from $r=0$.
Using these properties, it can be shown that the multiplication in Eq.~\eqref{diagonal-inverse} gives for
 the element in the top left corner of $\U$ the expression
\begin{align}\label{U11}
U_{11} &= \sum_{k=1}^M V_{1k} D_{kk} V_{k1}^\dagger = \sum_{k=1}^M |V_{1k}|^2 e^{i (2k-1-M)\theta} \notag\\
 &= \sum_{k=1}^M \binom {M-1}{k-1} \, |u^2|^{M-k} |v^2|^{k-1} e^{i (2k-1-M)\theta} \notag\\
 &= \sum_{k=1}^M \binom {M-1}{k-1} \, (|u^2|e^{-i\theta})^{M-k} (|v^2|e^{i\theta})^{k-1} \notag\\
 &= \left(|u^2|e^{-i\theta}+|v^2|e^{i\theta}\right)^{M-1} .
\end{align}
In a similar manner, we find 
\be\label{U1M}
U_{1M} = [uv (e^{i\theta}-e^{-i\theta})]^{\K-1}.
\ee

By comparing Eqs.~\eqref{U11} and \eqref{U1M} to Eqs.~\eqref{U-corners} we conclude that the parameters $u$ and $v$ are related to $a$ and $b$ as
\bse\label{ab(uv)}
\begin{align}
a &= |u|^2e^{-i\theta} + |v|^2 e^{i\theta} \label{atheta}, \\ 
b &= uv (e^{i\theta}-e^{-i\theta}) = 2iuv \sin\theta .
\end{align}
\ese
From here,
\bse\label{pq}
\begin{align}
|u|^2 &= \frac{\sin\theta-\Im\,a}{2\sin\theta}, \\
|v|^2 &= \frac{\sin\theta+\Im\,a}{2\sin\theta}, \\
uv &= \frac{-ib}{2\sin\theta} .
\end{align}
\ese

By using Eq.~\eqref{diagonal-inverse}, we find the multi-pass propagator,
\be\label{M^N}
\U^N = \V\D\V^\dagger\V\D\V^\dagger \cdots \V\D\V^\dagger
= \V\D^N\V^\dagger .
\ee
The $N$th power of the diagonal matrix $\D$ reads
\be\label{diagN}
\D^N =
\left[\begin{array}{ccccc}
   e^{-iN(M-1)\theta}&0 & \cdots & 0 \\
0&e^{-iN(M-3)\theta}& \cdots & 0\\
 \vdots& \vdots & \ddots & \vdots \\
  0 &0& \cdots& e^{iN(M-1)\theta}
\end{array}\right] .
\ee
Then, as in \eqref{M^N}, we obtain
\bse\label{aNbN(uv)}
\begin{align}
a_N &= |u|^2 e^{-iN\theta} + |v|^2 e^{iN\theta} , \\
b_N &= uv \left(e^{iN\theta}-e^{-iN\theta}\right) = 2iuv\sin(N\theta) .
\end{align}
\ese
By substituting Eq.~\eqref{pq} into Eq.~\eqref{aNbN(uv)} we find
\bse\label{aNbN-2}
\begin{align}
a_N &=\cos (N\theta) + i\Im(a) \dfrac{\sin(N\theta)}{\sin (\theta)},\\
b_N &=b \dfrac{\sin (N\theta)}{\sin (\theta)},
\end{align}
\ese
the same as Eqs.~\eqref{aNbN}.

The relations derived here between the propagator elements of the single-pass and $N$-pass interaction of the general $M$-state quantum system with the Majorana symmetry allow one to conduct two types of tasks: (i) given the action of the single interaction find the action of the repeated multiple interactions, and (ii) deduce the action of the single interaction by measuring the result of the multiple repetition of this interaction. 
These can be very useful in designing the best scenarios for coherent amplification of quantum gate errors and hence precise quantum gate tomography, as well as for enhanced quantum sensing of electric and magnetic fields by amplification of frequency shifts \cite{Vitanov2021}.
Indeed, having an exact analytic relation between the single-pass and multiple-pass processes enables the accurate determination of tiny gate errors or frequency shifts from the amplified error or shift.

We emphasize that the $M$-state Majorana system presents a clear benefit in this respect compared to the simple two-state system ($M=2$): the corner elements of the propagator [see Eqs.~\eqref{U-corners}] are the $(M-1)$-st power of the respective elements $a$ and $b$ of the two-state system.
If one of the parameters $a$ or $b$ is very small, then the $(M-1)$-st power will only make it smaller and closer to 0.
However the $(M-1)$-st power of the other parameter (which in modulus should be close to 1) will make it deviate much more strongly from 1 than for $M=2$. 
This deviation will be further amplified by using multiple interactions instead of a single one. 

Finally, we point out that in this second approach of derivation, we have used the explicit form of the diagonalizing matrix $\V$, which is composed of the eigenvectors of the Majorana propagator $\U$, with their elements obeying Eqs.~\eqref{pq}. 
This result can be useful by itself, e.g. for developing adiabatic control approaches \cite{Vitanov2001, Vitanov2017}, or the so-called ``shortcuts to adiabaticity'' \cite{STA}.

\subsection{Example: three-state system}

\begin{figure}[tb]
\includegraphics[width=0.90\columnwidth]{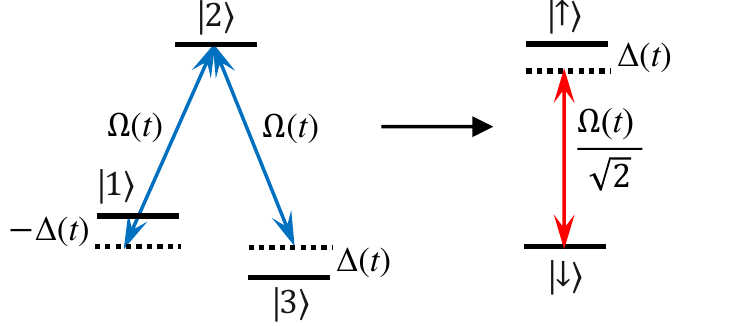}
\caption{
Majorana decomposition for the three-state $\Lambda$ system, with the equivalent two-state system.  
}
\label{fig:Maj1}
\end{figure}

The propagator $\U_{3}$ shown in \eqref{U3Maj} was constructed by the equation for the matrix elements $U_{\k\l}$ of Eq.~\eqref{Ukl}. 
The Hamiltonian is defined by Eq.~\eqref{HMajorana} and it is
\be
H_{3}(t) =
\left[\begin{array}{ccc}
-\Delta(t) & \frac{\Omega(t)}{\sqrt{2}}&0\\
\frac{\Omega^*(t)}{\sqrt{2}}&0&\frac{\Omega(t)}{\sqrt{2}}\\
0&\frac{\Omega^*(t)}{\sqrt{2}}&\Delta(t)
\end{array} \right],
\ee
The system is shown schematically in Fig.~\ref{fig:Maj1}. 
The multi-pass propagator $\U_{3}^N$ can be obtained from the single one $\U_{3}$ by the substitution $a\to a_N$ and $b \to b_N$ according to Eq.~\eqref{aNbN}, 
\be\label{U3MajN}
\U_{3}^N =
\left[\begin{array}{ccc}
 a_N^2 & \sqrt{2} a_N b_N & b_N^2 \\
 -\sqrt{2} a_N b_N^* & |a_N|^2-|b_N|^2 & \sqrt{2} b_N a_N^* \\
 b_N^{*2} & -\sqrt{2} a_N^* b_N^* & a_N^{*2} 
\end{array}
\right].
\ee
Therefore, already for $M=3$ states we have quadratic powers in the Cayley-Klein parameters and this amplification is further boosted by the application of $N$ interactions.

For example, if the Cayley-Klein parameter $a$ is real then $a^2=\cos^2\theta$ is the probability for no transition and $p_2= 1- a^2 =\sin^2\theta$ is the transition probability for a single-pass in the two-state system.
The $N$-pass transition probability reads
\be
p_2^{(N)} = p_2 \frac{\sin^{2} N\theta}{\sin^{2}\theta} 
= \sin^{2} N\theta.
\ee
Correspondingly, the transition probability in the three-state Majorana system from state 1 to state 3 is $p_3=\sin^4\theta$, and the $N$-pass transition probability reads
\be
p_3^{(N)} = p_3 \frac{\sin^{4} N\theta}{\sin^{4}\theta} 
= \sin^{4} N\theta.
\ee
A real $a$ occurs for resonant excitation ($\Delta=0$) and also when the Rabi frequency is a symmetric function of time while $\Delta$ is anti-symmetric \cite{Vitanov2021}.

The results in this example can be used for efficient tomography of qutrit gates as well as for quantum sensing with qutrits. Similar relations can be derived for larger number of states $M$.


\section{Systems with the Morris-Shore symmetry} \label{Sec:Morris-Shore}

\subsection{Single MS Propagator}

\begin{figure}[ht]
\includegraphics[width=0.90\columnwidth]{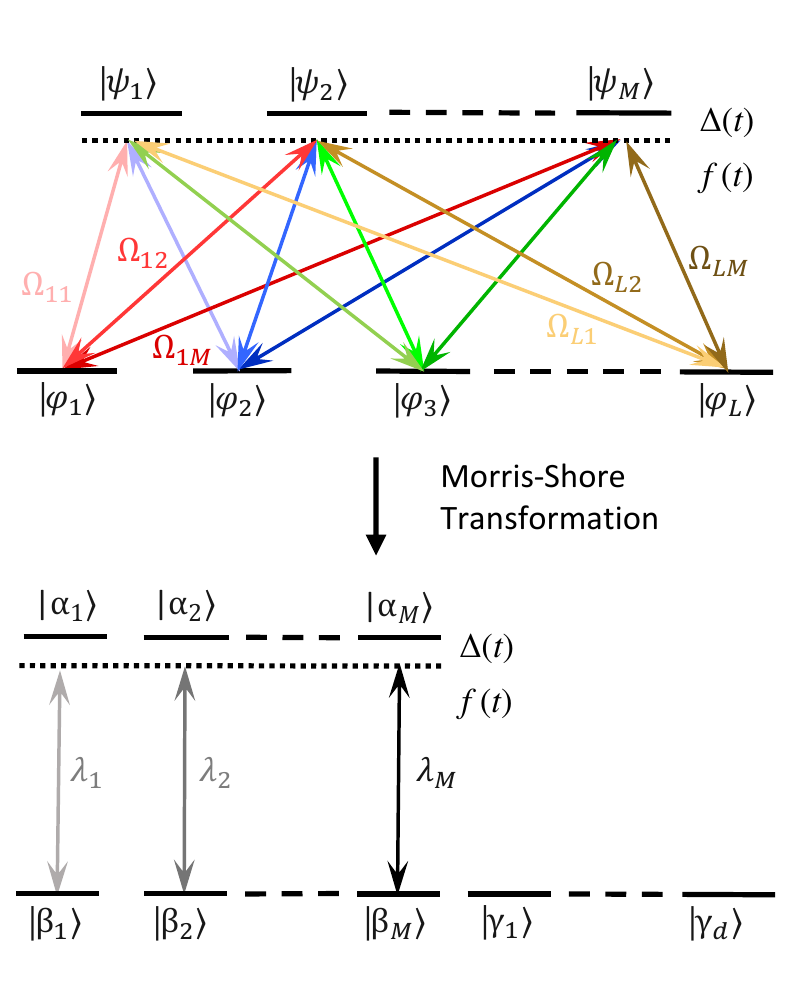}
\caption{
The Morris-Shore transformation. 
A multistate system consisting of two coupled sets of degenerate levels is transformed in to a set of $M$ independent two-state systems and a set of $d=L-M$ decoupled dark ground states. 
All couplings have same time dependence $f(t)$ and same detunings $\Delta(t)$ .  
}
\label{fig:MST1}
\end{figure}

The Morris-Shore transformation (MST) \cite {Morris1983,Rangelov2006,Kyoseva2006,Kyoseva2008,Zlatanov2020,Kis2002,Shore2013} is a powerful tool for reducing the dynamics of a certain class of multistate systems to the dynamics of one or more two-state systems.
The MS system is shown schematically in Fig.~\ref{fig:MST1}. 
It consists of two sets of states: a ground set with $L$ states and an exited set with $M$ states. 
There are couplings, quanttified by the Rabi frequencies $\Omega_{lm}f(t)$, only between states from different levels ($l=1,2,...,L$;  $m=1,2,...,M$), and $f(t)$ is a common time dependence of all couplings. %
All couplings have the same detunings $\Delta(t)$. 
The Hamiltonian of the MS system can be written as
\be
\H(t)=\frac{1}{2}\left[
\begin{array}{cc}
\mathbf{O_L} & f(t)\Omg\\
f(t)\Omg^\dagger&\Delta(t)\mathbf{1_M}  \\
\end{array}
\right], \quad\\
\ee
where the constant matrix $\Omg$ is $L\times M$ dimensional,
\be\label{OmegaMatrix}
\Omg=\left[
\begin{array}{cccc}
\Omega_{11} & \Omega_{12}&\cdots&\Omega_{1M}\\
\Omega_{21} & \Omega_{22}&\cdots&\Omega_{2M}\\
\cdots & \cdots&\ddots&\vdots\\
\Omega_{L1} & \Omega_{L2}&\cdots&\Omega_{LM}\\
\end{array}
\right],
\ee
and $\Omega_{lm}$ ($l=1,2,...,L;\quad m=1,2,...,M$) are arbitrary complex constants .
The MS transformation reduces the multi-state dynamics, which have a Hilbert space dimension of $L+M$ [Fig.~\ref{fig:MST1} (top)] to a set of $M$ independent two-state systems with additional $d=L-M$ decoupled (dark) states [Fig.~\ref{fig:MST1} (bottom)].
In the MS basis, the Hamiltonian has the block matrix form
\be\label{HMSt}
\wtl{\H}(t)=\S\H(t)\S^\dagger = 
\frac{1}{2}\left[\begin{array}{cc}
\mathbf{O_L} & f(t)\wtl{\Omg}\\
f(t)\wtl{\Omg}^\dagger&\Delta(t)\mathbf{1_M}  \\
\end{array}\right] ,
\ee
where $\S$ is a constant unitary matrix, defined by two square unitary matrices $\S_L$ and $\S_M$ with dimensions of $L$ and $M$, respectively,
\be\label{SS}
\S=\left[
\begin{array}{cc}
\S_L & \mathbf{O}\\
\mathbf{O} & \S_M  \\
\end{array}
\right],\quad \S\S^\dagger=\S^\dagger\S=\mathbf{1_{(L+M)}}.
\ee
Then, by using Eqs.~\eqref{HMSt} and \eqref{SS}, the transformed coupling matrix $\wtl{\Omg}$ can be expressed as
\be
\wtl{\Omg} = \S_L \Omg \S_M^\dagger
\ee

The matrices $\S_L$ and $\S_M$ are defined by the condition that they diagonalize $\Omg^\dagger\Omg$ and $\Omg\Omg^\dagger$, i.e.
\bse\label{AB}
\begin{align}
\S_L \Omg\Omg^\dagger \S_L^\dagger &= \mathbf{1_L}, \\
\S_M \Omg^\dagger\Omg \S_M^\dagger &= \mathbf{1_M}, 
\end{align}
\ese
and are found by solving these equations.
The $M$-dimensional square matrix $\Omg^\dagger\Omg$ has $M$ generally nonzero eigenvalues $\lambda^2_m$ $(m=1,2,...,M)$. 
The $L$-dimensional square matrix $\Omg\Omg^\dagger$ has the same $M$ eigenvalues as $\Omg^\dagger\Omg$ and additional $d=L-M$ zero eigenvalues, corresponding to the dark states.

The transformed Hamiltonian ~\eqref{HMSt} acquires the form
\be\label{HMS-matrix}
\wtl{\H}(t)=\left[\begin{array}{cc}
\mathbf{O}_{d \times d} & \mathbf{O}_{d \times 2M} \\
 \mathbf{O}_{2M \times d} & \wtl{\H}_c(t) 
\end{array}\right],
\ee
where $\wtl{\H}_c(t)$ is formed by 4 square $M \times M$ matrices,
\be\label{HMSc}
\wtl{\H}_c(t) = \left[\begin{array}{cc}
\mathbf{O} & {\bm \Lambda} f(t) \\
{\bm \Lambda} f(t) & \Delta(t) \mathbf{1}\\
\end{array}\right],
\ee
where 
\be
{\bm \Lambda} = \left[\begin{array}{cccc}
\lambda_1 & 0 & \cdots & 0 \\
0 & \lambda_2 & \cdots & 0\\
\vdots & \vdots & \ddots & \vdots \\
0 & 0 & \cdots & \lambda_M
\end{array}\right].
\ee
With an appropriate reordering of the states (i.e., the rows and the columns), described by a matrix $\R$, $\wtl{\H}_c(t)$ can be cast into the block matrix form $\R^{-1} \wtl{\H}_c(t) \R = \wtl{\H}_b(t)$, with
\be\label{HMS-2SS}
\wtl{\H}_b(t) = \left[\begin{array}{cccc}
\wtl{\H}_1(t) & \bm{0} & \cdots & \bm{0} \\
\bm{0} & \wtl{\H}_2(t) & \cdots & \bm{0} \\
\vdots & \vdots & \ddots & \vdots \\
\bm{0} & \bm{0} & \cdots & \wtl{\H}_M(t)
\end{array}\right],
\ee
with
\be
\wtl{\H}_m(t) = \left[\begin{array}{cc}
0 &\lambda_m f(t) \\
\lambda_m f(t) & \Delta(t)
\end{array}\right].
\ee
being the Hamiltonian describing the $m$-th independent two-state system in the MS basis.

Each of these $M$ independent two-state Hamiltonians
generates $M$ independent two-state propagators with $M$ different pairs of CK parameters $a_k$ and $b_k$,
\be\label{Udecomp}
\wtl{\U}_m=\left[\begin{array}{cc}
a_m & \b_m\\
-b^*_m e^{-i\delta} & a^*_m e^{-i\delta}
\end{array}\right],
\ee
where $|a_m|^2 +|b_m|^2=1$ and
\be
\delta=\int_{0}^{T} \Delta(t) \,dt 
\ee
is a common accumulated phase for all $M$ independent systems.
The propagator in the reordered MS basis has the same block matrix structure as $\wtl{\H}_b(t)$,
\be\label{HMS-b}
\wtl{\U}_b = \left[\begin{array}{cccc}
\wtl{\U}_1 & \bm{0} & \cdots & \bm{0} \\
\bm{0} & \wtl{\U}_2 & \cdots & \bm{0} \\
\vdots & \vdots & \ddots & \vdots \\
\bm{0} & \bm{0} & \cdots & \wtl{\U}_M
\end{array}\right].
\ee
After the reordering operation $\R \wtl{\U}_b \R^{-1} = \wtl{\U}_c$ we find  the propagator of the full system in MS basis in a block matrix form of four $M \times M$ square matrices,
\be\label{UMS-matrix}
\wtl{\U} = \left[\begin{array}{cc}
\mathbf{1}_{d \times d} & \mathbf{O}_{d \times 2M} \\
 \mathbf{O}_{2M \times d} & \wtl{\U}_c 
\end{array}\right],
\ee
where
\be\label{UMc}
\wtl{\U}_c = \left[\begin{array}{cc}
\A & \B \\
 -\B^* e^{-i\delta} & \A^* e^{-i\delta} 
\end{array}\right],
\ee
with
\be
{\A} = \left[\begin{array}{cccc}
a_1 & 0 & \cdots & 0 \\
0 & a_2 & \cdots & 0 \\
\vdots &\vdots & \ddots & \vdots \\
 0 & 0 & \cdots & a_M
\end{array}\right],\quad
{\B} = \left[\begin{array}{cccc}
b_1 & 0 & \cdots & 0 \\
0 & b_2 & \cdots & 0 \\
\vdots &\vdots & \ddots & \vdots \\
 0 & 0 & \cdots & b_M
\end{array}\right].
\ee
The original propagator obeys the same transformation as the original Hamiltonian,
\be\label{U-single}
\U=\S^\dagger\wtl{\U}\S.
\ee

\subsection{Multi-pass MS Propagator}

With the expression~\eqref{U-single} for the single propagator ${\U}$ at hand, we can easily find the $N$-pass propagator by taking the $N$-th power of ${\U}$,
\be\label{UNor}
\U^N = \S^\dagger\wtl{\U}\S 
\S^\dagger\wtl{\U}\S 
\cdots 
\S^\dagger\wtl{\U}\S
= \S^\dagger \wtl{\U}^N \S ,
\ee
where $\wtl{\U}^N$ is the transformed MS propagator. 
As the MS system is decomposed into $M$ independent two-state systems, then all of them have $M$ independent evolutions, governed by the respective propagators $\wtl{\U}_k$ of Eq.~\eqref{Udecomp}. 
Indeed, we have from Eq.~\eqref{UMS-matrix}
\be\label{UMS-matrix-N}
\wtl{\U}^N = \left[\begin{array}{cc}
\mathbf{1}_{d \times d} & \mathbf{O}_{d \times 2M} \\
 \mathbf{O}_{2M \times d} & \wtl{\U}_c^N 
\end{array}\right] .
\ee
We have $\wtl{\U}_c^N = \R\wtl{\U}_b^N \R^{-1}$,
and $\wtl{\U}_b^N$ is readily derived,
\be\label{HMS-b-N}
\wtl{\U}_b^N = \left[\begin{array}{cccc}
\wtl{\U}_1^N & \bm{0} & \cdots & \bm{0} \\
\bm{0} & \wtl{\U}_2^N & \cdots & \bm{0} \\
\vdots & \vdots & \ddots & \vdots \\
\bm{0} & \bm{0} & \cdots & \wtl{\U}_M^N
\end{array}\right].
\ee
We can find all $\wtl{\U}_m^N$ using Eqs.~\eqref{U2} and \eqref{U2N}, and then  construct the propagator of the full system.

In order to find $\wtl{\U}_m^N$, we first note that it does not possess the SU(2) symmetry as its determinant is $e^{-i\delta}$, and hence we cannot use Eqs.~\eqref{U2} and \eqref{U2N} directly because they require SU(2) symmetry.
Therefore, we represent $\wtl{\U}_m$ as
\be\label{U2V2}
\wtl{\U}_m = e^{-i\delta/2} \left[\begin{array}{cc} a_m e^{i\delta/2}  & b_m e^{i\delta/2} \\ -b_m^*e^{-i\delta/2} & a_m^*e^{-i\delta/2}   \end{array}\right],
\ee
where the matrix on the right-hand side is now SU(2) symmetric,
 which allows us to apply relations \eqref{U2} and \eqref{U2N}. 
We have
\be\label{UmN}
\wtl{\U}_m^N = e^{-iN\delta/2}
\left[\begin{array}{cc} a_m e^{i\delta/2}  & b_m e^{i\delta/2} \\ -b_m^*e^{-i\delta/2} & a_m^*e^{-i\delta/2} \end{array}\right]^N.
\ee
Then, by introducing the notation
\be\label{abdelta}
a_{m}' = a_m e^{i\delta/2}, \quad
b_{m}' = b_m e^{i\delta/2}, 
\ee
and using Eqs.~\eqref{U2} and \eqref{U2N}, we find 
\be\label{U2Nfin}
\wtl{\U}_m^N = \left[\begin{array}{cc}
a_{m N}'  & b_{m N}' \\ -b'^*_{m N} e^{-iN\delta} & a'^*_{m N}  e^{-iN\delta}
\end{array}\right],
\ee
where
\bse \label{DrelM}
\begin{align}
a_{m N}' &= \left[\cos (N\theta_{m}') + i\,\Im(a_{m}') \dfrac{\sin(N\theta_{m}')}{\sin (\theta_{m}')} \right] e^{-iN\delta/2}, \\
b_{m N}' &= b_{k}' \dfrac{\sin (N\theta_{m}')}{\sin (\theta_{m}')} e^{-iN\delta/2}, \\
\theta_{m}' &= \arccos (\Re\,a_{m}') .
\end{align}
\ese

The relations \eqref{DrelM}, together with ~\eqref{abdelta} give the connection between the single $\wtl{\U}_m$ and the repeated $\wtl{\U}_m^N$ propagators. 
Thereby we find the multi-pass propagator of the $M$ MS two-state systems [cf. Eq.~\eqref{UMS-matrix-N}],
\be\label{UcN}
\wtl{\U}_c^N = \left[\begin{array}{cc}
\A_N & \B_N \\
 -\B_N^* e^{-iN\delta} & \A_N^* e^{-iN\delta} 
\end{array}\right],
\ee
with
\bse
\begin{align}
\A_N &= \left[\begin{array}{cccc}
a_{1N}' & 0 & \cdots & 0 \\
0 & a_{2N}' & \cdots & 0 \\
\vdots &\vdots & \ddots & \vdots \\
 0 & 0 & \cdots & a_{MN}'
\end{array}\right],
\\
\B_N &= \left[\begin{array}{cccc}
b_{1N}' & 0 & \cdots & 0 \\
0 & b_{2N}' & \cdots & 0 \\
\vdots &\vdots & \ddots & \vdots \\
 0 & 0 & \cdots & b_{MN}'
\end{array}\right].
\end{align}
\ese
Next, we find $\wtl{\U}^N$ from Eqs.~\eqref{UMS-matrix-N} and \eqref{UcN}.
Finally, we find the original $N$-pass propagator $\U^N$ by the transformation ~\eqref{UNor}, ie $\U^N = \S^\dagger\wtl{\U}^N\S$.

Because $\S$ is a constant matrix and it appears in both $\U$ of Eq.~\eqref{U-single} and $\U^N$ of Eq.~\eqref{UNor} in the same manner, it follows that $\U^N$ can be obtained from $\U$ by the substitutions $a_m\to a_{m N}'$ and $b_m \to b_{m N}'$ and $e^{i\delta} \to e^{iN\delta}$, according to the connections \eqref{DrelM}.

As it has been noticed in ~\cite{Kyoseva2008}, in the interaction representation for a single propagator, one can get rid of the phase $\delta$ by removing the phase factor  $e^{-i\delta}$. 
This means that for single propagators, the phase $\delta$ can be considered as unimportant. 
Now we see that for the multi-pass MS propagator we could also remove the phase factor $e^{-iN\delta}$ by switching to the interaction representation, so that only $a'_{m N}$ and $b'_{m N}$ will remain. 
However, these CK parameters depend on $\delta$ according to \eqref{DrelM}. 
Which means, that for repeated MS propagators the phase $\delta$ becomes already important.

\subsection{Special cases: multipod systems}

\subsubsection{Single-pass multipod propagator}

\begin{figure}[tb]
\includegraphics[width=0.90\columnwidth]{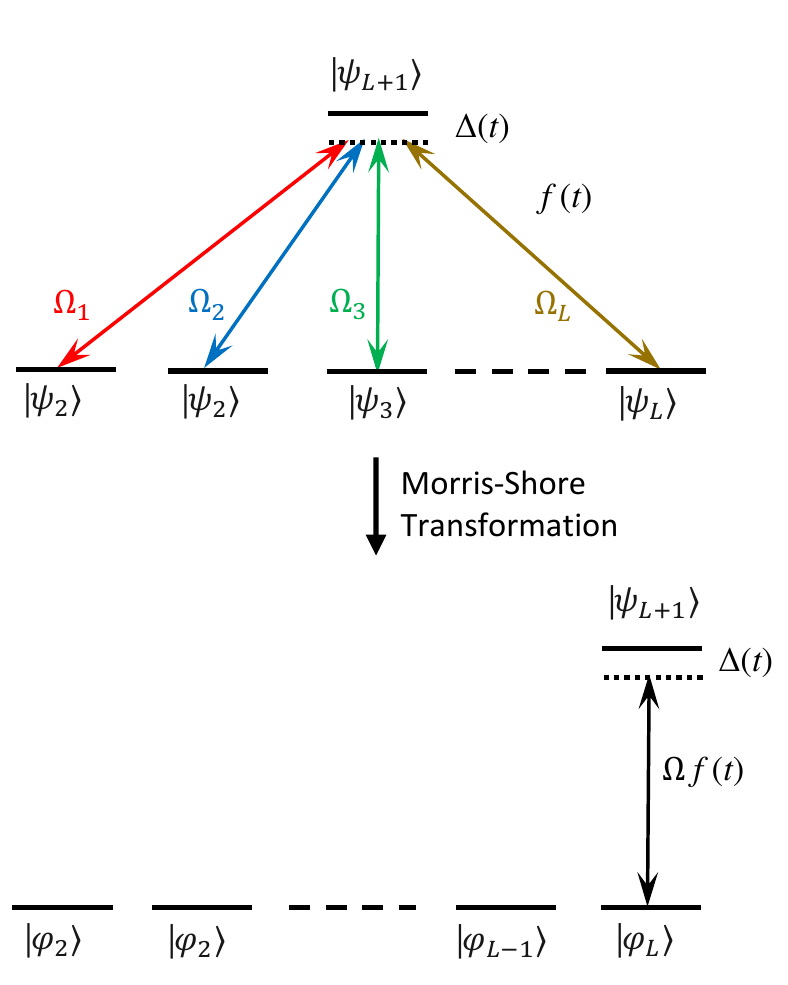}
\caption{
The Morris-Shore transformation for the multipod system, which consists of $L$ degenerate ground states and a single excited state. 
The system is transformed in to a single two-state system and a set of $L-1$ decoupled (dark) ground states. 
All couplings have the same time dependence $f(t)$ and the same detuning $\Delta(t)$.
}
\label{fig:MST2}
\end{figure}

As the first example, we consider the multipod system, shown schematically in Fig.~\ref{fig:MST2}.  
It consists of $L$ ground states and a single excited state, $M=1$. 
All ground states are coupled to the excited state but not between themselves.
All couplings have the same time dependence $f(t)$ and the same detuning $\Delta(t)$, but their magnitudes can be all different.
Therefore, its Hamiltonian fulfils the MS symmetry. 


The constant matrix $\Omg$ of Eq.~\eqref{OmegaMatrix} has the dimension of $L \times 1$ and the Hamiltonian reads
\be\label{Hk}
\H = \frac12 \left[\begin{array}{ccccc}
0 & 0 & \cdots & 0 & \Omega_1f(t) \\
0 & 0 & \cdots & 0 & \Omega_2f(t) \\
\vdots & \vdots & \ddots & \vdots & \vdots \\
0 & 0 & \cdots & 0 & \Omega_L f(t) \\
\Omega^*_1f(t) & \Omega^*_2f(t) & \cdots & \Omega^*_L f(t) & 2\Delta(t)
\end{array}\right],
\ee
where $\Omega_l$ $(l=1,2,...,L)$ are arbitrary complex constants.
The MS transformation from the original to the MS basis for the Hamiltonian and the propagator reads
\be \label{HMSeq}
\wtl{\H} = \S^\dagger \H \S , \quad
\wtl{\U} = \S^\dagger \U \S ,
\ee
where the constant transformation matrix $\S$ has the form 
\be\label{S_Npod}
\renewcommand{\arraystretch}{2}
{\S} = \left[\begin{array}{ccccccc}
\frac{\Omega_2^*}{X_2}&\frac{\Omega_1\Omega_3^*}{X_2X_3} &\frac{\Omega_1\Omega_4^*}{X_3X_4}& \cdots &\frac{\Omega_1\Omega_L^*}{X_{L-1}X_L}&\frac{\Omega_1}{X_L} &0\\
-\frac{\Omega_1^*}{X_2}&\frac{\Omega_2\Omega_3^*}{X_2X_3} &\frac{\Omega_2\Omega_4^*}{X_3X_4}& \cdots &\frac{\Omega_2\Omega_L^*}{X_{L-1}X_L}&\frac{\Omega_2}{X_L}&0\\
0&-\frac{X_2}{X_3}&\frac{\Omega_3\Omega_4^*}{X_3X_4}&\cdots &\frac{\Omega_3\Omega_L^*}{X_{L-1}X_L}&\frac{\Omega_3}{X_L}&0\\
0&0&-\frac{X_3}{X_4}&\cdots&\frac{\Omega_4\Omega_L^*}{X_{L-1}X_L}&\frac{\Omega_3}{X_L}&0\\
 \vdots& \vdots & \vdots & \ddots&\vdots&\vdots&\vdots \\
0&0&\cdots &0&-\frac{X_{L-1}}{X_L}&\frac{\Omega_L}{X_L}&0\\
0 &0& \cdots& 0& 0&0&1
\end{array}\right].
\ee
The real constants $X_l$ are given by
\be
X_l=\sqrt{\sum\nolimits_{k=1}^{l}|\Omega_k|^2},\quad (l=2,3,...,L).
\ee
The MS transformation of Eq.~\eqref{HMSeq}, along with Eqs.~\eqref{Hk} and  ~\eqref{S_Npod}, gives the transformed MS Hamiltonian $\wtl{\H}(t)$, which reduces to an effective two-state system,
\be\label{HMS}
\wtl{\H} = \frac{1}{2}\left[\begin{array}{ccccc}
 0&0 & \cdots & 0&0 \\
0&0& \cdots & 0&0\\
 \vdots& \vdots & \ddots &\vdots&\vdots \\
0&0&\cdots &0&\Omega f(t)\\
0&0& \cdots& \Omega f(t)&2\Delta(t)
\end{array}\right],
\ee
with 
\be
\Omega = \sqrt{\sum\nolimits_{l=1}^{L}|\Omega_l|^2} .
\ee

The MS Hamiltonian \eqref{HMS} generates the propagator
\be\label{UMSwtl}
\wtl{\U} = \left[ \begin{array}{cccccc}
  1&0 & \cdots &0& 0&0 \\
0&1& \cdots &0& 0&0\\
 \vdots& \vdots & \ddots & \vdots&\vdots&\vdots \\
0&0&\cdots &1&0&0\\
0&0&\cdots &0&a&b\\
0 &0& \cdots& 0& -b^*e^{-i\delta}&a^*e^{-i\delta}
\end{array}\right],
\ee
which has only a pair of CK parameters $a,b$, because all other states in the MS basis are decoupled.
The original propagator is found by the inverse of Eq.~\eqref{HMSeq}, i.e. $\U = \S \wtl{\U} \S^\dagger $, 
or explicitly,
\be\label{UK}
\U \!=\! \left[ \! \begin{array}{cc}
    {\mathbf {1_L}} + (a-1) \dfrac{\ket{\Omega}\bra{\Omega}}{\Omega^2} & 
    \begin{matrix} 
    b\frac{\Omega_1}{\Omega} \\
    b\frac{\Omega_2}{\Omega} \\
    \vdots \\
    b\frac{\Omega_L}{\Omega}
    \end{matrix} \\ 
    \begin{matrix} 
    -b^*\frac{\Omega_1^*}{\Omega} e^{-i\delta}  & -b^*\frac{\Omega_2^*}{\Omega} e^{-i\delta} & \cdots & -b^*\frac{\Omega_L^*}{\Omega} e^{-i\delta} \end{matrix} & a^* e^{-i\delta}
    \end{array} \!\right]\!,
\ee
where $\ket{\Omega} = \{\Omega_1,\Omega_2,\ldots,\Omega_L\}^T$ and hence $\bra{\Omega} = \{\Omega_1,\Omega_2,\ldots,\Omega_L\}^*$.

\subsubsection{Multi-pass multipod propagator}

According to Eq.~\eqref{Udecomp}, the $N$-pass propagator in the MS basis is
\be\label{UMSwtlN}
\wtl{\U}^N = \left[\begin{array}{cccccc}
  1&0 & \cdots &0& 0&0 \\
0&1& \cdots &0& 0&0\\
 \vdots& \vdots & \ddots & \vdots&\vdots&\vdots \\
0&0&\cdots &1&0&0\\
0&0&\cdots &0& a'_{N} & b'_{N}\\
0 &0& \cdots& 0& -b'^*_{N} e^{-iN\delta} & a'^*_{N} e^{-iN\delta}
\end{array}\right],
\ee
where the repeated CK parameters $a'_{N}$ and $b'_{N}$ are determined with the connections \eqref{DrelM} by the single parameters $a$, $b$ and $\delta$ (i.e. without the subscript $k$).
The multi-pass propagator $\wtl{\U}^N$ has the same form as the single-pass one  $\wtl{\U}$ \eqref{UMSwtl}. Therefore we use again the expression for the original propagator \eqref{UK}, by making the substitutions $a\to a'_{N}$,\quad $b \to b'_{N}$ and $e^{-i\delta} \to e^{-iN\delta}$. Thereby we find
\be\label{U-multipod}
\U = \left[\begin{array}{cc}
    {\mathbf {1_L}} + (a'_N-1) \dfrac{\ket{\Omega}\bra{\Omega}}{\Omega^2} & 
    \begin{matrix} 
    b'_N\frac{\Omega_1}{\Omega} \\
    b'_N\frac{\Omega_2}{\Omega} \\
    \vdots \\
    b'_N\frac{\Omega_L}{\Omega}
    \end{matrix} \\ 
    \begin{matrix} 
    -b'^*_{N}\frac{\Omega_1^*}{\Omega} e^{-iN\delta} & 
    \cdots & -b'^*_{N}\frac{\Omega_L^*}{\Omega} e^{-iN\delta} \end{matrix} & a'^*_{N} e^{-iN\delta}
    \end{array}\right] .
\ee

As examples of the general multipod system, we consider explicitly the multi-pass propagators for two important cases: the $\Lambda$ and tripod systems.

\subsubsection{Multipass propagator for the $\Lambda$ system}

\begin{figure}[tb]
\includegraphics[width=0.90\columnwidth]{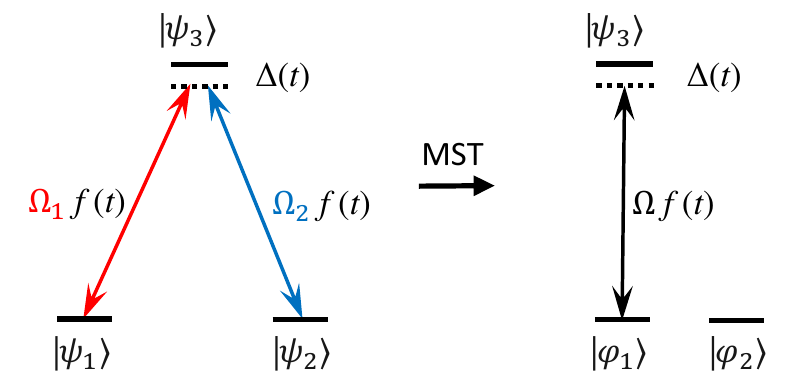}
\caption{
Morris-Shore transformation for the $\Lambda$ system.  
}
\label{fig:MS_Lambda}
\end{figure}

The $\Lambda$ system is shown in Fig.~\ref{fig:MS_Lambda}. 
It is the most ubiquitous multistate system used, for instance, in stimulated Raman adiabatic passage (STIRAP) \cite{Vitanov2017}, as a Raman qubit (formed by the two lower states) \cite{Leibfried2003,Gaebler2016}, and a qutrit \cite{Ringbauer2018, Low2020, Ringbauer2021} in quantum information, etc.
There is a single condition for the applicability of the MS transformation: the two couplings must share the same time dependence $f(t)$, as shown in the figure.
The MS transformation matrix is
\be\label{S_Lambda}
\S_{\Lambda}=\left[
\begin{array}{ccc}
\frac{\Omega_2^*}{\Omega}&\frac{\Omega_1}{\Omega}&0\\
-\frac{\Omega_1^*}{\Omega}&\frac{\Omega_2}{\Omega}&0\\
0&0&1
\end{array}
\right],
\ee
and the multi-pass propagator reads
\be\label{ULambda}
\renewcommand{\arraystretch}{2}
\U_{\Lambda}^N \!=\!\! \left[\!\begin{array}{ccc}
  1+(a'_{N}-1)\frac{|\Omega_1|^2}{\Omega^2}&(a'_{N}-1)\frac{\Omega_1 \Omega_2^*}{\Omega^2}&b'_{N}\frac{\Omega_1}{\Omega} \\
(a'_{N}-1) \frac{\Omega_1^* \Omega_2}{\Omega^2} & 1+(a'_{N}-1) \frac{|\Omega_2|^2}{\Omega^2} & b'_{N} \frac{\Omega_2}{\Omega} \\
-b'^*_{N} \frac{\Omega_1^*}{\Omega} e^{-iN\delta} & -b'^*_{N} \frac{\Omega_2^*}{\Omega}e^{-iN\delta} & a'^*_{N} e^{-iN\delta}
\end{array}\!\right]\!,
\ee
where $\Omega=\sqrt{|\Omega_1|^2+|\Omega_2|^2}$.\\

\subsubsection{Multipass propagator for the tripod system}

\begin{figure}[tbph]
\includegraphics[width=0.90\columnwidth]{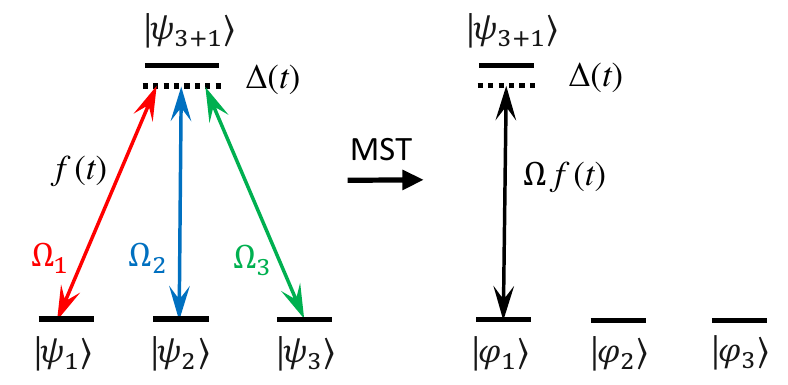}
\caption{
Morris-Shore transformation for the tripod system.  
}
\label{fig:MS_3pod}
\end{figure}

The tripod system, shown in Fig.~\ref{fig:MS_3pod}, is another popular system because it possesses two dark states, which can serve as a decoherence-free qubit in topologic quantum information \cite{Unanyan1998, Unanyan1999, Unanyan2004}.
It also allows a great flexibility in the creation of arbitrary coherent superposition of the three lower states, and can be used more generally as a very convenient implementation of a qutrit.

The MS transformation matrix is
\be\label{S_3pod}
\renewcommand{\arraystretch}{2}
\S_{T} = \left[\begin{array}{cccc}
\frac{\Omega_2^*}{X_2}&\frac{\Omega_1\Omega_3^*}{X_{2}X_3}&\frac{\Omega_1}{X_3} &0\\
-\frac{\Omega_1^*}{X_2}&\frac{\Omega_2\Omega_3^*}{X_{2}X_3}&\frac{\Omega_2}{X_3}&0\\
0&-\frac{X_2}{X_3}&\frac{\Omega_3}{X_3}&0\\
0&0&0&1
\end{array}\right],
\ee
and the multipass propagator is
\bwt
\be\label{U3pod}
\renewcommand{\arraystretch}{2}
\U_{T}^N = \left[\begin{array}{cccc}
  1+(a'_N-1)\frac{|\Omega_1|^2}{\Omega^2}&(a'_N-1)\frac{\Omega_1 \Omega_2^*}{\Omega^2}&(a'_N-1)\frac{\Omega_1 \Omega_3^*}{\Omega^2} &b'_N\frac{\Omega_1}{\Omega}\\
(a'_N-1)\frac{\Omega_1^* \Omega_2}{\Omega^2} & 1+(a'_N-1)\frac{|\Omega_2|^2}{\Omega^2} &(a'_N-1)\frac{\Omega_2 \Omega_3^*}{\Omega^2} &b'_N\frac{\Omega_2}{\Omega}\\
(a'_N-1)\frac{\Omega_1^* \Omega_L}{\Omega^2} &(a'_N-1)\frac{\Omega_2^* \Omega_L}{\Omega^2}&1+(a'_N-1)\frac{|\Omega_3|^2}{\Omega^2} &b'_N\frac{\Omega_3}{\Omega}\\
-b'^*_{N} \frac{\Omega_1^*}{\Omega}e^{-iN\delta} & -b'^*_{N} \frac{\Omega_2^*}{\Omega}e^{-iN\delta} & b'^*_{N} \frac{\Omega_3^*}{\Omega}e^{-i N\delta} & a'^*_{N} e^{-iN\delta}
\end{array}\right],
\ee
\ewt
where $\Omega=\sqrt{|\Omega_1|^2+|\Omega_2|^2+|\Omega_3|^2}$.

\section{Conclusion\label{Sec:Conclusion}}

In this paper, we have derived explicit analytic formulae describing the interaction of multistate quantum systems with either the Majorana SU(2) or the Morris-Shore symmetry with a driving field consisting of $N$ identical single-step fields.
For a single-step interaction the dynamics of these systems can be reduced to the dynamics of one or more two-state systems.
We have used this feature in order to derive the propagators for these two types of systems in terms of the parameters of the two-state propagators and the number of interactions $N$.
Because of the availability of the propagators, the results allows one to readily find out the state of the quantum system for arbitrary initial conditions.

Our results can find applications in the development of quantum control method for multistate systems by using the well known methods for two-state systems.
In particular, one can find exact analytic solutions for the multipass dynamics of multistate systems using the well known single-pass two-state analytic models. 

On the other hand, the explicit analytic formulae make it possible to develop precise methods for quantum gate tomography of multistate systems (e.g. qutrits, and qudits in general, as well as an ensemble of a few qubits) by repeated application of the quantum gate, which quickly amplifies its infidelity to levels which can be measured very accurately \cite{Vitanov2020, Vitanov2021}.
Then the analytic connections between the single-pass and multi-pass propagator parameters allow one to deduce the single-pass ones from the multi-pass ones.

\acknowledgements

We dedicate this paper to the memory of our dear friend Bruce W. Shore, who revealed to us the power of the beautiful Morris-Shore transformation, one of the most elegant approaches to understanding the complex dynamics of multistate quantum systems.  

This work is supported by the European Commission's Quantum Flagship Project 820314 (MicroQC).

\end{document}